\documentclass[reprint,amsmath,amssymb,aps,prl,superscriptaddress]{revtex4-1}

\usepackage{graphicx}% Include figure files
\usepackage{dcolumn}% Align table columns on decimal point
\usepackage{bm}% bold math
\usepackage{xcolor}

\usepackage{subcaption}
\newcommand{\labelphantom}[1]{%
{\phantomsubcaption%
\label{#1}}%
}%

\begin{document}

\title{First Access to ELM-free Negative Triangularity at Low Aspect Ratio}
%Decorrelation of Second Stability Access and ELM-Free Operation in Negative Triangularity Discharges on MAST-U

\author{A.O. Nelson}
\email[Corresponding author: ]{a.o.nelson@columbia.edu}
\affiliation{Columbia University, New York City, New York, USA}

\author{C. Vincent}
\affiliation{UKAEA-CCFE, Culham Science Centre, Abingdon, Oxon, UK}

\author{H. Anand}
\affiliation{General Atomics, San Diego, California, USA}

\author{J. Lovell}
\affiliation{Oak Ridge National Laboratory, Oak Ridge, Tennessee, USA}

\author{J.F. Parisi}
\affiliation{Princeton Plasma Physics Laboratory, Princeton, New Jersey, USA}

\author{H. S. Wilson}
\affiliation{Columbia University, New York City, New York, USA}

\author{K. Imada}
\affiliation{York Plasma Institute, Heslington, York, UK}

\author{W.P. Wehner}
\affiliation{General Atomics, San Diego, California, USA}

% \author{\color{red}J. Barr}
% \affiliation{General Atomics, San Diego, California, USA}

\author{M. Kochan}
\affiliation{UKAEA-CCFE, Culham Science Centre, Abingdon, Oxon, UK}

\author{S. Blackmore}
\affiliation{UKAEA-CCFE, Culham Science Centre, Abingdon, Oxon, UK}

\author{G. McArdle}
\affiliation{UKAEA-CCFE, Culham Science Centre, Abingdon, Oxon, UK}

% \author{\color{red}A. Lvovskiy}
% \affiliation{General Atomics, San Diego, California, USA}

% \author{\color{red}V. Soukhanovskii} 
% \affiliation{Lawrence Livermore National Laboratory, Livermore, California, United States of America}

% \author{\color{red}O. Sauter}
% \affiliation{Ecole Polytechnique Fédérale de Lausanne, Swiss Plasma Center, Lausanne, Switzerland}

% \author{\color{red}J. McClenaghan}
% \affiliation{General Atomics, San Diego, California, USA}

\author{S. Guizzo}
\affiliation{Columbia University, New York City, New York, USA}

\author{L. Rondini}
\affiliation{Columbia University, New York City, New York, USA}

\author{S. Freiberger}
\affiliation{Columbia University, New York City, New York, USA}

\author{C. Paz-Soldan}
\affiliation{Columbia University, New York City, New York, USA}

\author{the MAST-U Team}

\begin{abstract}
A plasma scenario with negative triangularity (NT) shaping is achieved on MAST-U for the first time. While edge localized modes (ELMs) are eventually suppressed as the triangularity is decreased below $\delta\lesssim-0.06$, an extended period of H-mode operation with Type-III ELMs is sustained at less negative $\delta$ even through access to the second stability region for ideal ballooning modes is closed. This documents a qualitative difference from the ELM-free access conditions documented in NT scenarios on conventional aspect ratio machines. The electron temperature at the pedestal top drops across the transition to ELM-free operation, but a steady rise in core temperature as $\delta$ is decreased allows for similar normalized $\beta$ in the ELM-free NT and H-mode positive triangularity shapes.
\end{abstract}
\date{\today}

\maketitle

%%%%%%%%%%%%%%%%%%%%%%%%%%%%%%%%%%%%%%%%%%%%%%%%%%%%%%%%%%%%%%%
%%%%%%%%%%%%%%%%%%% SECTION 1: INTRODUCITON %%%%%%%%%%%%%%%%%%%
%%%%%%%%%%%%%%%%%%%%%%%%%%%%%%%%%%%%%%%%%%%%%%%%%%%%%%%%%%%%%%%

\section{Introduction}

Tokamak scenarios exploring negative triangularly (NT) shaping, where the plasma x-points are situated at a larger radius than the magnetic axis, have recently gathered significant interest from the international community due to their favorable confinement and power-handling properties \cite{marinoni_brief_2021}. In machines with conventional aspect ratios $A=R_\mathrm{0}/a_\mathrm{minor}\sim2.5-4$, where $R_\mathrm{0}$ and $a_\mathrm{minor}$ are respectively the major and minor radii, NT experiments are often characterized by reduced electron heat flux throughout their volumes compared to similar discharges with positive triangularity (PT), which comes from a reduction in core turbulent transport \cite{Camenen2007, Austin2019}. This leads to an enhancement of the core pressure gradient in NT plasmas that improves confinement over matched PT discharges \cite{paz-soldan_simultaneous_2024, coda_enhanced_2022}. Notably, these same plasmas do not induce edge localized modes (ELMs), even at high heating powers \cite{nelson_robust_2023}. This ELM-free behavior allows NT fusion power plant designs to avoid issues associated with plasma-wall interactions during ELM events \cite{Medvedev2015, nelson_prospects_2022, rutherford_manta_2024}. The larger major radius of the x-point in NT scenarios also provides additional room for divertor heat flux spreading, highlighting the potential power-handling benefits of NT plasmas \cite{Kikuchi2019, miller_power_2024}.

While NT scenarios show tangible promise as an power plant candidate in conventional aspect ratio designs, they have not yet been fully explored in spherical tokamaks (STs) with lower aspect ratio ($A\lesssim2$). Initial gyrokinetic simulations of NT scenarios at low $A$, both with \texttt{gs2} \cite{davies_kinetic_2022} and \texttt{GENE} \cite{balestri_physical_2024}, predict that the confinement enhancement over PT discharges observed at conventional aspect ratio might be lost in NT ST scenarios, particularly in regimes dominated by trapped electron modes. In addition, NT plasmas are predicted to be strongly unstable to long-wavelength kinetic ballooning modes (KBMs) across the plasma in high-$\beta$, reactor-relevant STs \cite{davies_kinetic_2022}. This results from a closure of access to the so called “second stability” region for ballooning modes, potentially limiting the maximum achievable $\beta$ for NT scenarios in STs \cite{Medvedev2015}. 

Notably, the loss of second stability access has also been strongly correlated with the prevention of ELMs in conventional aspect ratio NT discharges \cite{Medvedev2015, saarelma_ballooning_2021, nelson_robust_2023}. Access to second stability has long been associated with large H-mode pedestals \cite{bishop_stability_1986, sabbagh_transition_1989, nelson_prospects_2022}, so it is initially unsurprising that closing the stability window for ideal ballooning modes would prevent H-mode access. On DIII-D, second stability access can be closed by directly scanning the triangularity $\delta$ below some critical value ($\delta_\mathrm{crit,\,DIII-D}\sim-0.15$) while holding other plasma and machine parameters (such as the injected power, plasma current, toroidal magnetic field and density) constant \cite{nelson_robust_2023}. This shape change modifies the local shear near the plasma edge, closing access to the second stability regime and preventing the formation of an ELMy H-mode regime by severely limiting the allowable edge pressure gradient \cite{saarelma_ballooning_2021}. Negative triangularity plasmas on TCV follow a similar pattern, where closure of the second stability region is encountered in ELM-free NT plasmas \cite{Medvedev2015, merle_pedestal_2017, coda_enhanced_2022}.

As noted in the literature, however, closure of the second stability region is not itself always sufficient for H-mode prevention in PT plasmas, particularly at lower aspect ratios \cite{dickinson_towards_2011, nelson_prospects_2022, parisi_kinetic-ballooning-limited_2024}. Further, experimentally measured gradients from DIII-D ELM-free plasmas with $\delta\sim-0.5$ are not always indicative of being directly limited by the ideal ballooning mode \cite{nelson_robust_2023}. These two facts suggest (1) that closure of the second ballooning stability region alone may not be sufficient to prevent ELMs, especially in regimes STs where first-stable H-modes are more common, and (2) that additional transport mechanisms that further limit the pressure gradient below the ballooning limit may be present even in conventional aspect ratio machines, especially when $\delta\ll0$. 

In this Letter, we present initial observations and modeling from the first ELM-free NT discharge achieved on MAST-U. Importantly, this work demonstrates a distinct decorrelation between the closure of the second stability region and the prevention of ELMy H-mode operation that has not yet been observed in conventional aspect ratio machines. Section~\ref{sec:methods} details the access conditions for NT operation on MAST-U, leading to documentation of the stability modeling in section~\ref{sec:stability}. In section~\ref{sec:conclusion}, the potential consequences of this result are briefly discussed, particularly with respect to the construction of a predictive model for the NT pedestal height, which may ultimately set the performance of NT reactor scenarios.

%%%%%%%%%%%%%%%%%%%
%%% SECTION ONE %%%
%%%%%%%%%%%%%%%%%%%

\section{NT Operation on the MAST-U Spherical Tokamak}
\label{sec:methods}

\begin{figure}
    \includegraphics[width=1\linewidth]{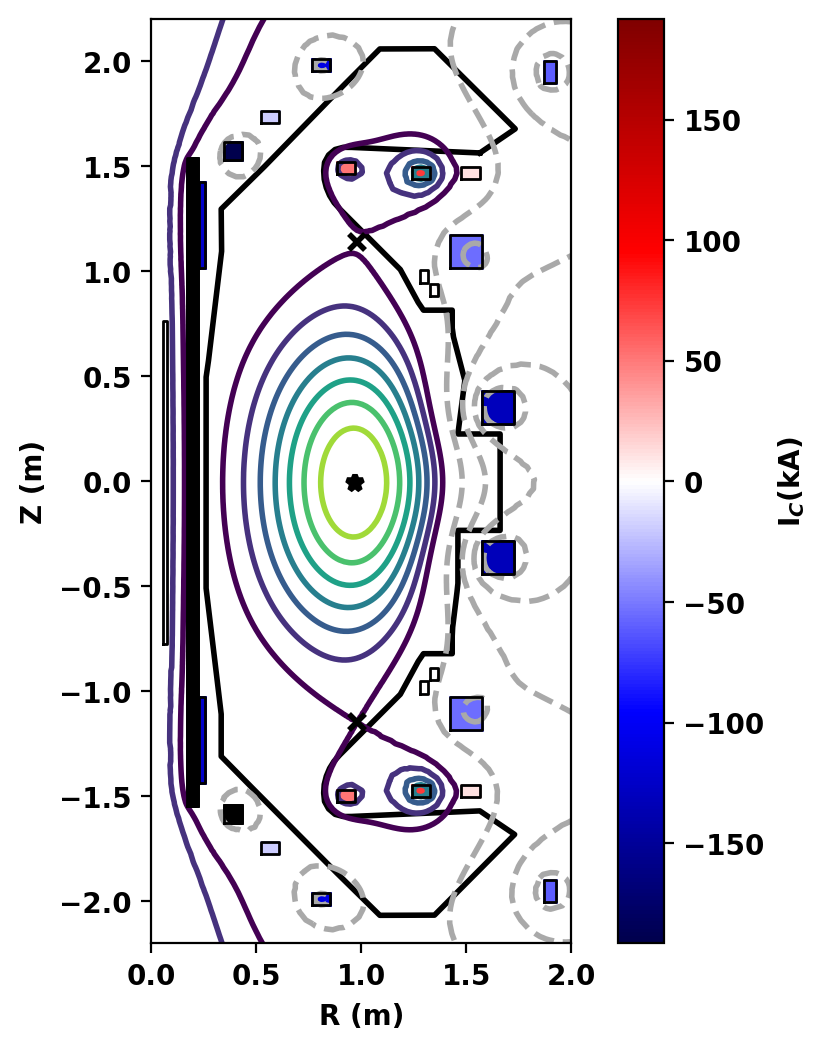}
    \caption{MAST-U discharge \#49463 at $800\,$ms is diverted, nearly up-down symmetric, and has $\delta<0$.}
    \label{fig:eq}
\end{figure}

The magnetic equilibrium for MAST-U discharge \#49463, which features diverted negative triangularity operation% near the most negative $\delta$ that can be achieved within the MAST-U vacuum vessel
, is reconstructed by the \texttt{TokaMaker} code \cite{hansen_tokamaker_2024} in figure~\ref{fig:eq}. This discharge was achieved during a snowflake divertor development experiment in which the \texttt{GA-TokSys} code \cite{humphreys_development_2007} was employed to reduce the plasma squareness at high elongation \cite{mcardle_mast_2020, anand_real-time_2024}. During this particular discharge, the lower x-point (subject to vertical position control) left the x-point control region as the plasma elongation was reduced, at which point the plasma control system continued to push the plasma in feedback. Luckily, this resulted in a smooth triangularity scan at constant power and near constant density, providing an excellent first NT discharge for the initial physics investigations presented in this work. Further details on this control scheme, including techniques that can be employed to create stable NT discharges on MAST-U, will be the subject of a subsequent publication.

\begin{figure}
    \includegraphics[width=1\linewidth]{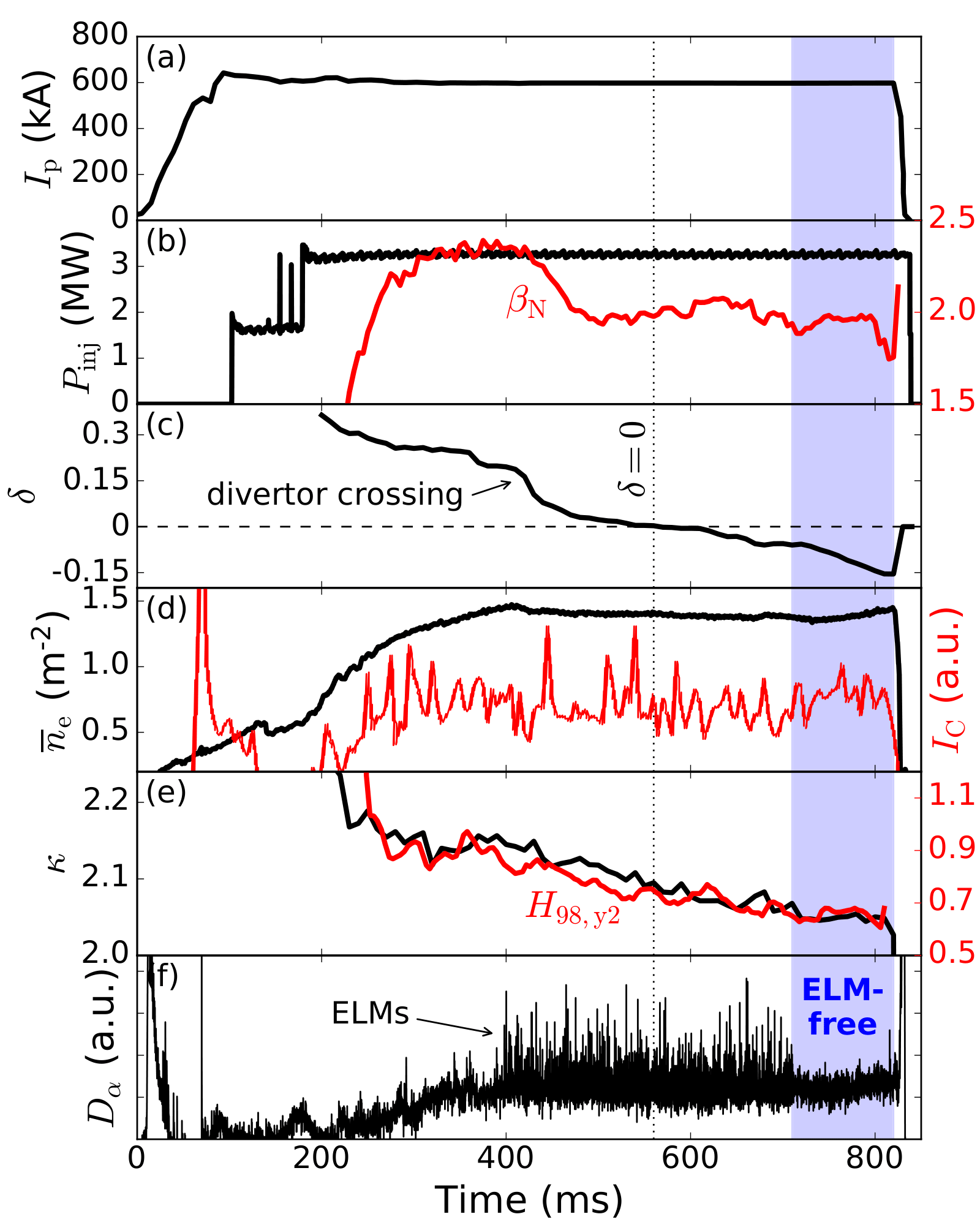}
    \labelphantom{fig:traces-a}
    \labelphantom{fig:traces-b}
    \labelphantom{fig:traces-c}    
    \labelphantom{fig:traces-d}    
    \labelphantom{fig:traces-e}
    \labelphantom{fig:traces-f}
    \caption{(a) The plasma current $I_\mathrm{p}$, (b) injected beam power $P_\mathrm{inj}$ (black) and $\beta_\mathrm{N}$ (red), (c) triangularity $\delta$, (d) line-integrated average density $\overline{n}_\mathrm{e}$ (black) and carbon intensity $I_\mathrm{C}$ (red) from SPRED, (e) the plasma elongation $\kappa$ (black) and normalized confinement $H_\mathrm{98,y2}$ (red), and (f) $D_\mathrm{\alpha}$ emission lines for MAST-U discharge \#49463. The ELM-free period is marked in blue.}
    \label{fig:traces}
\end{figure}

Figure~\ref{fig:traces} shows major time traces for this discharge, highlighting the stable nature of the plasma as the shape is changed. The discharge has constant plasma current ($I_\mathrm{p}=600\,$kA) and beam heating power ($P_\mathrm{inj} = 3\,$MW). A shape scan starts around $t=200\,$ms, as can be seen in the $\delta$ scan presented in figure~\ref{fig:traces-c}. We note that the strong dip in $\delta$ near $t=420\,$ms is a result of the inner strike-line moving on to the outboard portion of the vacuum vessel.  The line-integrated average density (figure~\ref{fig:traces-d}) grows until $t=400\,$ms, at which point an ELMy H-mode appears. The density then remains roughly constant for the remainder of the discharge, although some slight rise in the density can be observed after the transition to the ELM-free state, presumably linked to reduced pumping in the absence of ELMs (as also occurs before H-mode is achieved.) 
%A closer look at the $D_\mathrm{\alpha}$ emission line around the transition time is provided in figure~\ref{fig:dalpha}, where it can be seen that the last spike (which occurs around $t\sim815\,$ms) in the $D_\mathrm{\alpha}$ is not an ELM. Inspection of equilibria reconstructions reveal that the plasma boundary hits the vessel wall near this time, leading to a total loss of shape control around $t\sim830\,$ms. 

Evidence of the ELMs can be seen in the $D_\mathrm{\alpha}$ traces presented in figure~\ref{fig:traces-f}. Notably, the plasma remains in H-mode with rapid Type-III ELMs as the triangularity drops below $\delta=0$ until a transition to an ELM-free state is made at around $t_\mathrm{crit}\sim710\,$ms when $\delta_\mathrm{crit}\lesssim-0.06$.  
The observation of ELM-free behavior below a critical triangularity $\delta_\mathrm{crit}$ is qualitatively consistent with expectations concerning the behavior of NT scenarios from the DIII-D and TCV tokamaks \cite{nelson_robust_2023, coda_enhanced_2022}. However, it remains to be conclusively demonstrated that this phenomena is a robust property of MAST-U NT discharges. Interestingly, infinite-$n$ stability modeling suggests that access to this ELM-free scenario may be governed by different physics than in DIII-D or TCV, as is discussed in section~\ref{sec:stability}. 

Over the course of discharge \#49463, the normalized confinement $H_\mathrm{98y2}$, as calculated with the \texttt{TRANSP} code \cite{Poli2018}, decreases steadily with decreasing $\kappa$ (figure~\ref{fig:traces-e}) from $H_\mathrm{98y2}\sim0.9$ at $300\,$ms to $H_\mathrm{98y2}\sim0.66$ during the ELM-free phase. No significant changes in $H_\mathrm{98y2}$ are observed near $\delta=0$ or $\delta=\delta_\mathrm{crit}$. The normalized pressure $\beta_\mathrm{N}$ (figure~\ref{fig:traces-b}) drops significantly from $\beta_\mathrm{N}\sim2.4$ to $\beta_\mathrm{N}\sim2$ when the inner strikeline moves across the divertor opening but remains constant at $\beta_\mathrm{N}\sim2$ for the rest of the discharge. Notably, little change in the impurity content is observed across this discharge, as indicated by the carbon emission measurements captured with the SPRED diagnostic shown in figure~\ref{fig:traces-d} \cite{fonck_multichannel_1982}. Slight ($\sim10-20\%$) increases in the line-averaged density and the C and He SPRED emissions are observed during the ELM-free phase, potentially suggestive of reduced impurity transport under ELM-free conditions. 

% \begin{figure}
%     \includegraphics[width=1\linewidth]{fig03_dalpha.png}
%     \caption{A zoom-in of the $D_\mathrm{\alpha}$ traces from figure~\ref{fig:traces-e} around the transition time. The ELM-free period is marked in blue.}
%     \label{fig:dalpha}
% \end{figure}

\begin{figure}
    \includegraphics[width=1\linewidth]{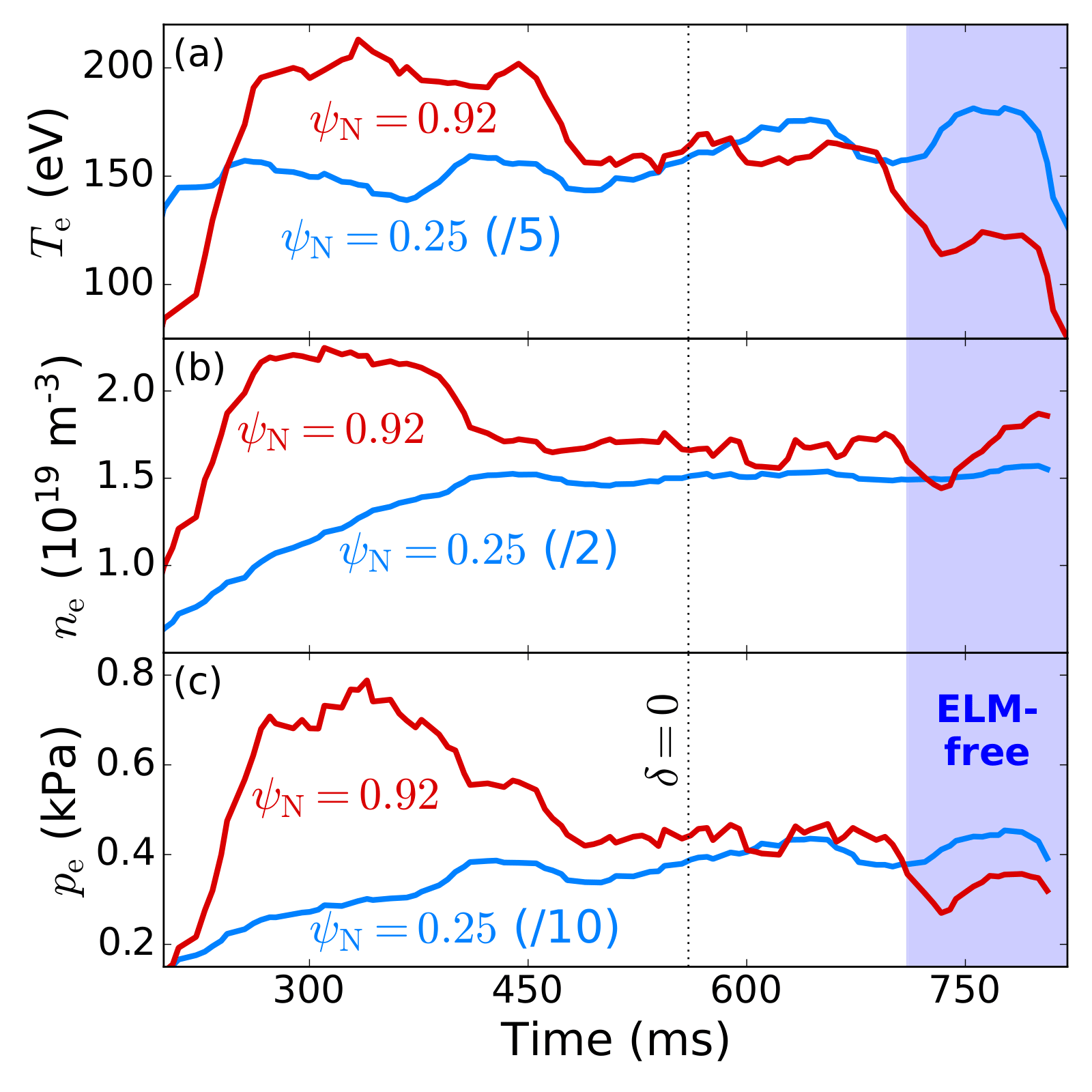}
    \labelphantom{fig:profs-a}
    \labelphantom{fig:profs-b}
    \labelphantom{fig:profs-c}
    \caption{The (a) electron temperature, (b) electron density and (c) electron pressure as a function of time for two radial locations: the core (blue) and the pedestal top (red). Note that the core profiles are scaled down by constant factors for visibility.}
    \label{fig:profs}
\end{figure}

Representative traces of the electron temperature ($T_\mathrm{e}$), density ($n_\mathrm{e}$) and pressure ($p_\mathrm{e}$) at the plasma core ($\psi_\mathrm{N}\sim0.25$) and near the pedestal top ($\psi_\mathrm{N}\sim0.92$) are shown in figure~\ref{fig:profs}. Near the pedestal top, $T_\mathrm{e,ped}$ drops significantly both when the inner strikeline transitions to the outboard side of the vessel around $t\sim420\,$ms and then again when the ELM-free state is achieved at $t\sim710\,$ms. The edge density $n_\mathrm{e,ped}$ stays roughly constant as the strikeline shifts, but drops upon the transition to the ELM-free state before growing again towards the end of the discharge. Notably, the core traces display markedly different trends: while $n_\mathrm{e,core}$ remains roughly constant after H-mode is achieved, $T_\mathrm{e,core}$ actually rises steadily as $\delta$ is decreased, reaching its highest values in the ELM-free portion of the discharge when $\delta$ is the most negative. This behavior is reproduced qualitatively in initial simulations with the \texttt{STEP} code \cite{lyons_flexible_2023}, which recovers an increase in the core $T_\mathrm{e}$ as $\delta$ is decreased at fixed density and injected power that results from a reduced \texttt{TGLF/TGYRO} SAT-2 \cite{staebler_theory-based_2007} electron heat flux at more negative triangularity. The rise in core temperature at constant density translates to a steady rise in the core pressure as $\delta$ is decreased, as seen in figure~\ref{fig:profs-c}. Similar qualitative behavior demonstrating a decrease in edge pressure but a retention of high core pressures has been observed in NT plasmas on DIII-D \cite{nelson_robust_2023, paz-soldan_simultaneous_2024} and TCV \cite{Camenen2007, Sauter2014} and is often associated with a reduction in the electron thermal conductivity in NT scenarios.

Further characterization of the plasma performance (including transport mechanisms, $\beta_\mathrm{N}$ limits and confinement scalings) as a function of $\delta<0$ requires additional investigations over a dedicated set of NT discharges on MAST-U and are thus not considered here. Exploration of the heat-flux width, ELM behavior and shaping limits will also be included in future studies.

\section{Decorrelation of Second Stability Closure and ELM-free NT Operation}
\label{sec:stability}

On the DIII-D and TCV tokamaks, which have conventional aspect ratios, access to ELM-free conditions via NT shaping occurs suddenly at a critical triangularity ($\delta_\mathrm{crit}$). This phenomena has been strongly correlated with the closure of the second stability region for infinite-$n$ ballooning modes, which introduces an additional transport mechanism at the plasma edge, preventing the growth of steep pressure gradients in the NT edge \cite{saarelma_ballooning_2021, nelson_prospects_2022, nelson_robust_2023, merle_pedestal_2017, nelson_characterization_2024}. However, as mentioned in \cite{nelson_prospects_2022, nelson_characterization_2024}, it is in general not necessarily sufficient in tokamak plasmas that second stability access be closed to ensure L-mode operation. Broadly speaking, if the collisionality is high enough to suppress the bootstrap current, local shear can remain high enough to keep the edge in the first stable region even when weak peeling-ballooning modes are destabilized. This effect is particularly evident in devices with low aspect ratio, and it is not uncommon for MAST-U H-mode discharges with positive triangularity to form pedestals that reside in the first-stable region \cite{dickinson_towards_2011}. On NSTX, the prevalence of first-stable H-modes can be observed in the bifurcation of pedestal widths into `wide' and `narrow' branches, which are in turn set by variations in the stability of kinetic ballooning modes \cite{parisi_kinetic-ballooning-bifurcation_2024, parisi_kinetic-ballooning-limited_2024}. 

\begin{figure}
    \includegraphics[width=1\linewidth]{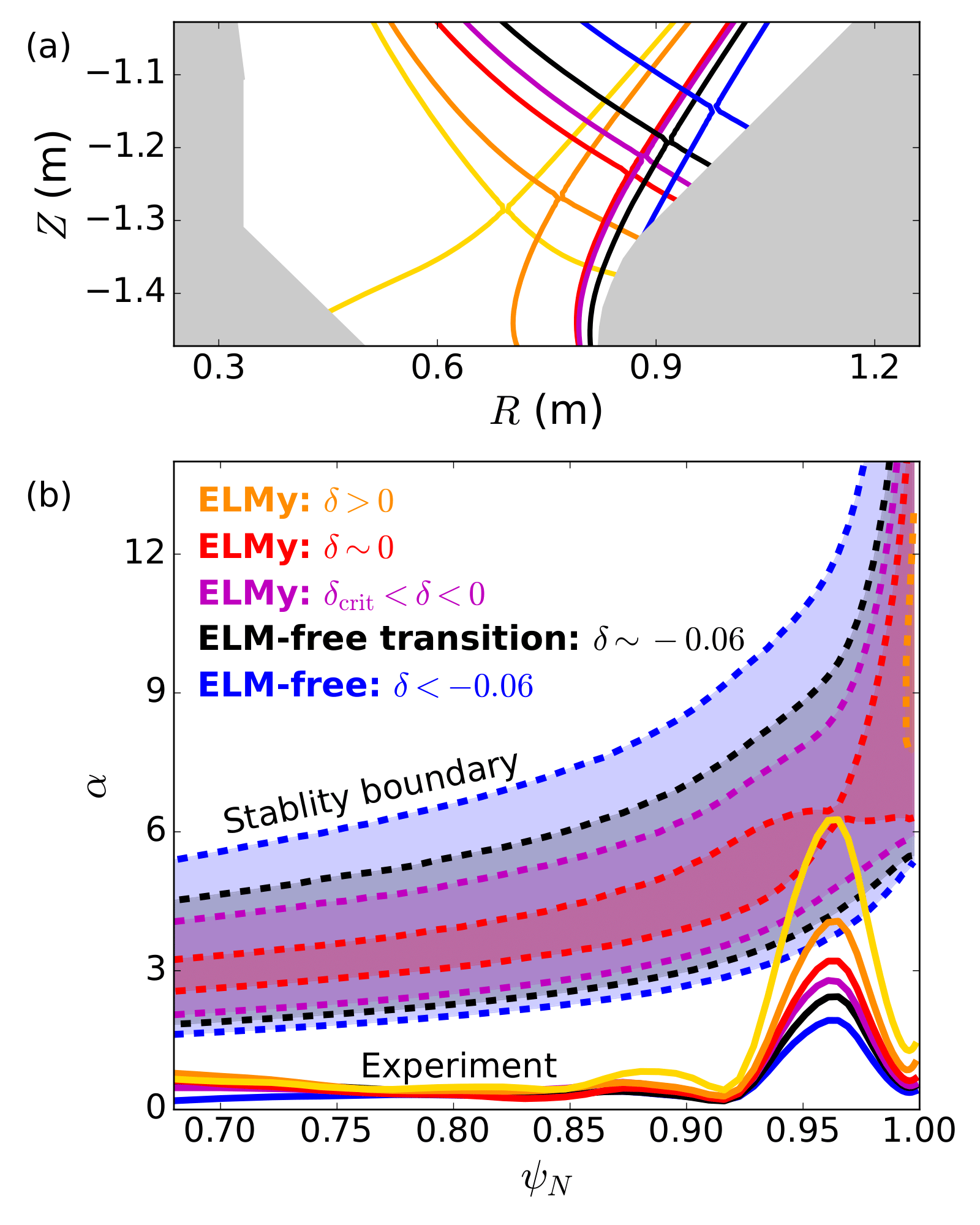}
    \labelphantom{fig:baloo-a}
    \labelphantom{fig:baloo-b}
    \caption{(a) A zoom-in of the divertor region and (b) the corresponding infinite-$n$ ballooning stability for a series of times during MAST-U discharge \#49463. ELMy timeslices are colored yellow, orange, red and magenta. The stability region (dashed lines) is open originally but closes well in advance of the ELM-free transition (black). The final ELM-free state is shown in blue. At no point during the evolution of this discharge is the experimental normalized pressure gradient (solid lines) directly limited by the first stability limit for infinite-$n$ modes.}
    \label{fig:baloo}
\end{figure}

The ideal ballooning stability of tokamak discharges can be easily assessed using the \texttt{BALOO} code \cite{miller_power_2024}, which requires as input well-converged kinetic equilibrium reconstructions. For this work, kinetic equilibria were generated every $10\,$ms using the kineticEFITtime workflow provided in the OMFIT framework \cite{Meneghini2015}. This workflow allows for iteration between profile fits to the electron and ion profiles and a free-boundary EFIT reconstruction \cite{Lao1985}. The stability results from this study are presented in figure~\ref{fig:baloo}, which shows the normalized pressure gradient ($\alpha$) corresponding to the stability boundaries (dashed lines) and experimental conditions (solid lines) for five key times throughout discharge \#49463. 

In figure~\ref{fig:baloo}, the first four times (yellow -- $300\,$ms, orange -- $400\,$ms, magenta -- $480\,$ms and red -- $590\,$ms) are all from periods of the discharge when the plasma is firmly in a Type-III ELMy regime. No stability boundary is found by the \texttt{BALOO} code during the earliest times ($t\lesssim400\,$ms), allowing the pressure gradient at the edge to grow to high values. However, the stability boundary rapidly closes around $t\sim480\,$ms when the average triangularity measured at the last closed flux surface (LCFS) nears zero. Throughout the evolution of this discharge, the peak pressure gradient in the edge region decreases as $\delta$ is decreased. ELMy H-mode behavior is robustly sustained until a triangularity of $\delta\sim-0.06$ is reached around $t=710\,$ms (black lines in figure~\ref{fig:baloo}). An ELM-free time slice ($t=800\,$ms) is shown in blue for comparison. Notably, max($\alpha$) remains similar for the entire ELM-free phase ($t=720-830\,$ms) and does not continue to drop as $\delta$ is further decreased during this time.

Strikingly, this analysis reveals a decorrelation between the closure of ballooning stability window and the prevention of ELMs by NT shaping. While previously suggested as a possible complication to the no-ELM NT scenario \cite{nelson_prospects_2022, qin_effect_2023, nelson_characterization_2024}, this has not yet been observed on either the DIII-D or TCV tokamaks, where closure of second stability typically coincides with the loss of H-mode in NT \cite{medvedev_beta_2008, merle_pedestal_2017, saarelma_ballooning_2021, nelson_characterization_2024}. 

To compensate for any uncertainties in the equilibrium reconstructions that could impact this result, over 500 variations of the kinetic equilibrium representing the plasma at $t=500\,$ms were generated by solving the fixed-boundary using the \texttt{CHEASE} code \cite{lutjens_chease_1996} with arbitrary profile modifications, following the workflow established in \cite{nelson_prospects_2022}. Scaling the edge pressure gradient, collisionality and current separately by factors of five was not sufficient to open access to the second stability region in this shape, suggesting that the conclusion that this MAST-U discharge achieves a first-stable H-mode regime at low values of $\delta<0$ is not an artifact of any workflow. 

As seen in figure~\ref{fig:traces-f}, an ELM-free state is still achieved in this low aspect ratio discharge, suggesting that additional physics may be needed to fully explain the transition to ELM-free states via NT shaping over a wider variety of plasma conditions. As the plasma conditions ($\overline{n}_\mathrm{e}$, $P_\mathrm{inj}$ and impurity content) were held constant over this scan, it is likely that the transition to an ELM-free state in discharge \#49463 is still largely a result of the shape changes. Interestingly, and as a result of the unique shapes achieved in this discharge, while the triangularity measured at the last closed flux surface drops below zero around $t=590\,$ms, the triangularity just inside of the LCFS (up to $\psi_\mathrm{N}\sim0.9$) remains positive until the ELM-free transition at $t=710\,$ms. This may suggest that the effects of NT must extend over a significant portion of the edge region to inhibit the transition to H-mode, which is likely true in conventional aspect ratio scenarios as well. 

\begin{figure}
    \includegraphics[width=1\linewidth]{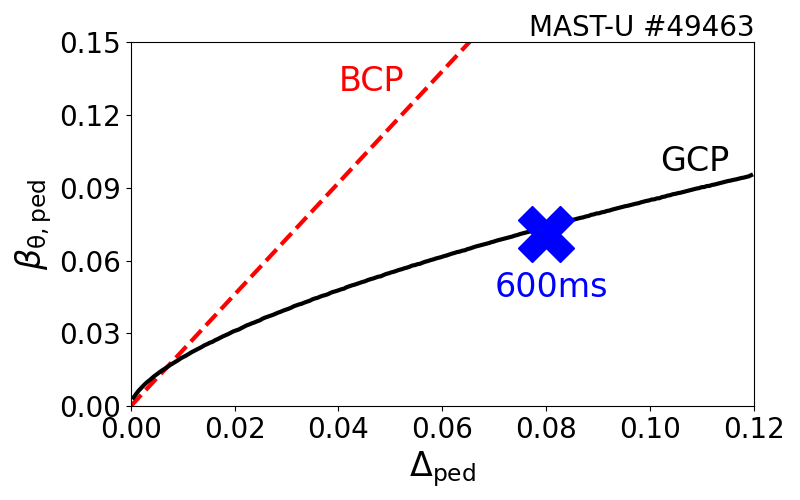}
    \caption{Calculations of the ballooning-critical pedestal (BCP; red, dashed) and gyrokinetic-critical pedestal (GCP; blacked, solid) for MAST-U discharge \#49463 at $600\,$ms, which is ELMy at modest $\delta<0$ with closed access to the second stability region.}
    \label{fig:gkped}
\end{figure}

To begin to assess possible mechanisms for pedestal height reduction responsible for the ELM-free behavior observed at strong NT on MAST-U, the \texttt{gk\_ped} code \cite{parisi_gk_ped_2023} was utilized to determine the Gyrokinetic Critical Pedestal (GCP) in discharge \#49463. This model combines linear gyrokinetics with self-consistent pedestal equilibrium variation to explicitly determine stability limits stemming from the Kinetic Ballooning Mode (KBM) in the plasma edge and has recently been utilized to reproduce experimental measurements of the pedestal width-height scalings in the NSTX tokamak, which also has low-aspect-ratio \cite{parisi_kinetic-ballooning-limited_2024}. This GCP is calculated for an ELMing case with $\delta>0$ ($600\,$ms) in figure~\ref{fig:gkped}, showing excellent agreement with the experimentally measured pedestal height ($\beta_\mathrm{\theta,ped}$) and width ($\Delta_\mathrm{ped}$), which are calculated from a modified-tanh fit to the edge pressure profile. Notably, the Ballooning Critical Pedestal (BCP), which is calculated directly from the infinite-$n$ ballooning stability boundary, lies well above this curve, in agreement with the \texttt{BALOO} modeling presented in figure~\ref{fig:baloo}. 

This result carries several important implications. First, it offers insight into the Type-III ELMy nature of these discharges. Initial investigations with the \texttt{ELITE} code \cite{Snyder2002} suggest that these plasmas sit well within the nominal stability region for Type-I ELMs -- a reduction in the diamagnetic stabilization factor by a factor of $\sim1/10$ from typical MAST-U values is needed to localize the experimental measurements along the predicted stability boundary. If the plasma is indeed limited by a KBM constraint that sits below the peeling-ballooning boundary, as is suggested by \texttt{gk\_ped}, it would offer an justification for the reduced pedestal height and Type-III ELMy operation. Second, the prevalence of this KBM limit lends some credibility to the theoretical arguments made in \cite{davies_kinetic_2022}, which finds that the KBM will be a primary source of transport in ST NT scenarios with reactor-relevant parameters. Further study of the role of KBMs in setting the $\beta$ limits for NT scenarios on STs should be the subject of future work.

Finally, the reduction of the modeled GCP limit from the BCP provides a potential explanation for the decorrelation of second stability closure and ELM-free NT operation that distinguishes this regime from counterparts on conventional aspect ratio tokamaks. While the mechanisms for ELM suppression are still unclear, reduction of the pedestal height below the first stability limit for infinite-$n$ ballooning modes occurs already in the ELMy phase of this discharge as a result of the KBM limit. While potentially more applicable in low-aspect ratio devices \cite{parisi_kinetic-ballooning-bifurcation_2024, parisi_kinetic-ballooning-limited_2024}, such a reduction due to kinetic effects could also be present in conventional aspect ratio devices and offer a potential avenue for gradient reduction below the ideal ballooning limit, as observed in \cite{nelson_robust_2023, nelson_characterization_2024}. Future work, including similar KBM analysis on conventional aspect ratio machines and further study of the ELM-free NT regime on MAST-U across a broader parameter range, is needed to fully explore the validity of this model.

\section{Discussion and Conclusion}
\label{sec:conclusion}

This Letter reports on the first NT plasma achieved in a spherical tokamak. While further experiments, including on other STs like NSTX-U and SMART, are needed to fully characterize the performance and stability of this regime, several interesting observations can already be made due to the fortuitously stable plasma conditions achieved in this discharge. First, the qualitative behavior expected from NT operation on other machines concerning a sharp transition to an ELM-free regime without significant degradation of plasma performance is achieved. While $H_\mathrm{98y2}$ decreases across the shape scan with decreasing $\kappa$, no obvious decrease in $\beta_\mathrm{N}$ is associated with the transition to a shape with $\delta<0$ or when achieving H-mode inhibition. This optimistically suggests that NT operation of low aspect ratio devices follow the intuition developed on larger aspect ratio devices, though gyrokinetic predictions do predict degradation of plasma performance at very low $A$ \cite{davies_kinetic_2022, balestri_physical_2024}. Appropriate characterization of the confinement properties of the plasmas demands additional experimental investigation.

Perhaps more interestingly, at less negative triangularities an extended period of Type-III ELMing H-mode persists even though second stability for ideal ballooning modes is closed. The decoupling of second stability closure and access to the NT ELM-free state has not yet been observed on either DIII-D or TCV, which typically transition directly from a Type-I ELMy H-mode into the NT ELM-free state when the ballooning stability bifurcates. Access to first-stable ELMy H-mode at small absolute values of $\delta<0$ could have significant consequences on the feasibility of this approach as a pilot plant scenario for low aspect ratio machines. This supports arguments presented in reference~\cite{nelson_prospects_2022} regarding consideration of the scalings presented therein only as bounding values on the required shaping parameters for access to the ELM-free state. If first-stable H-modes with Type-III ELMy can be achieved in a given configuration, more negative values of $\delta$ may be needed to achieve robust H-mode inhibition. 

We note further that this key result -- that additional physics mechanisms beyond the ideal ballooning stability likely impact the edge gradient limits in NT plasmas -- is not necessarily inconsistent with results obtained on other machines at larger aspect ratio \cite{nelson_characterization_2024}. On DIII-D, while access to ELM-free conditions is strongly correlated with the closure of ideal ballooning stability \cite{saarelma_ballooning_2021}, a majority of ELM-free NT discharges sustain gradients well below the maximum gradient allowed by first stability limit for ballooning modes. This can be observed clearly in figure 3 of reference~\cite{nelson_robust_2023}, which shows numerous ELM-free DIII-D equilibria sitting well under the first stability limit. Taken together, these results are indicative of additional transport mechanisms present in the NT edge that can both limit the edge gradient and robustly prevent ELMs. Future work on this subject should aim to identify this physics to enable reliable predictions of the NT edge under the conditions expected in future fusion power plants.

\section*{Acknowledgments}

Some calculations performed in this study were completed through the OMFIT framework \cite{Meneghini2015}, and the generated data are available upon request. This material is based upon work supported by the U.S. Department of Energy, Office of Science, under Awards DE-SC0022272, DE-SC0018991, DE-AC05-00OR22725 and DE-FC02-04ER54698. This work has been funded in part by the EPSRC Energy Programme [grant number EP/W006839/1]. To obtain further information on the data and models underlying this paper please contact PublicationsManager@ukaea.uk.

% \bibliography{2024_MAST-U_NT}

\end{document}